\documentclass[twocolumn,showpacs,amsmath,amssymb]{revtex4}
\usepackage{graphicx}

\begin{document}

\title{Charge-impurity-induced Majorana fermions in topological superconductors}

\author{Tao Zhou}
\email{tzhou@nuaa.edu.cn}
\affiliation{College of Science, Nanjing University of Aeronautics and Astronautics, Nanjing 210016, China}

\author{Xiao-Jing Li}
\affiliation{College of Science, Nanjing University of Aeronautics and Astronautics, Nanjing 210016, China}

\author{Yi Gao}
\affiliation{Department of Physics and Institute of Theoretical Physics,
Nanjing Normal University, Nanjing, 210023, China}

\author{Z. D. Wang}
\email{zwang@hku.hk}
\affiliation{Department of Physics and Center of Theoretical and Computational Physics, The University of Hong Kong, Pokfulam Road, Hong Kong, China}

\date{\today}
\begin{abstract}
We study numerically Majorana fermions (MFs) induced by a charged impurity in topological superconductors. It is revealed from the relevant Bogoliubov-de
Gennes equations that (i) for quasi-one dimensional systems, a pair of MFs are bounded at the two sides of one charge impurity and well separated; and (ii) for a two dimensional square lattice, the charged-impurity-induced MFs are similar to the known pair of vortex-induced MFs, in which
one MF is bounded by the impurity while the other appears at the boundary. Moreover, the corresponding local density of states is explored,  demonstrating that the presence of MF states may be tested experimentally.

\end{abstract}
\pacs{71.10.Pm, 03.67.Lx, 74.90.+n}
\maketitle

\section{Introduction}
Topological superconductors have recently attracted tremendous attention  due to their exotic properties and potential applications~\cite{qi}. There are a number of candidate systems for realizing topological superconductors. One well-known example is the chiral $p+ip$ superconductor~\cite{read},
while its experimental realization is of great challenge.
A more realistic model
may be the realized in the spin-orbital coupled $s$-wave superconducting systems, such as the Cu$_x$Be$_2$Se$_3$ material
~\cite{hor,wra,andr,sasa}. Besides,  
topological superconductors may also be fabricated in heterostructure systems including a semiconductor with the spin-orbital interaction and a conventional $s$-wave superconductor~\cite{lutc,oreg,wil,rok,den,das,mou,jhlee}.
Another promising  candidate is the topological superfluid which may be also simulated by cold atom systems, noting that  both the $s$-wave pairing and  spin-orbital coupling have been implemented in cold atoms~\cite{bou,jkchin,lin,wang,cheu}.

As is known, one of the most prominent features of a topological superconductor lies in that topologically non-trivial Majorana fermions (MFs) may emerge in the system, which is relevant to the non-Abelian statistics and has potential applications in topological quantum computation~\cite{naya}.
It was
indicated that the MFs may exist in the vortex cores of a chiral $p+ip$ superconductors~\cite{read} and they should obey non-Abelian statistics~\cite{iva}. Later there are a number of theoretical proposals for realizing and probing the MFs in various topological superconducting systems~\cite{lfu,sato,zhcw,jliu,lind,sau,zhu,huhui,jiang,zhou, yxzhao}. Although significant experimental efforts have been made on MFs~\cite{wil,rok,den,das,mou,jhlee,nadj}, an unambiguous evidence for the MFs and a direct demonstration for their non-Abelian statistics are still waited.

On the other hand, the impurity effect has been an important issue in the study of unconventional superconductivity~\cite{bra}. For a $d$-wave superconducting system, one remarkable result is the existence of the mid-gap bound state near the impurity site, while such kind of bound state does not exist for the conventional $s$-wave superconductors.
For a topological superconducting system, the existence of the mid-gap state is of great interest and may be related to the MF modes. Thus the impurity effect is also of interest, which has been paid attention~\cite{jsau,xjliu,nagai,hhu,nag,naga}.
Different from the impurity effect of a topologically trivial $s$-wave superconductor, it has been reported that the in-gap bound states may be induced by a non-magnetic impurity~\cite{jsau,xjliu,nagai,hhu,nag,naga}.
The mid-gap state exists in the pure one dimensional system~\cite{xjliu}, or in the two dimensional system but considering a typical line-type potential~\cite{nagai}. While generally for the two and three dimensional systems,
the in-gap bound states appear at the finite energy and no zero-energy states exist near the impurity~\cite{nagai,hhu,nag}. As a result, no MFs are actually induced by a single impurity for two or three dimensional system.
 Notably, the previous studies of the single impurity effect focus merely on the single neutral non-magnetic in-plane impurity, in which
 the impurity term was treated theoretically as an additional potential on the impurity site. In contrast to them, we here investigate the effect of an off-plane charged impurity in a topological superconductor with the impurity term being simulated by an additional Coulomb potential,
 noting that the charge-impurity effect has intensively been studied for some condensed matter systems, such as high-T$_c$ superconductors~\cite{zqwang} and graphene layers~\cite{jhchen}.
More interestingly, we find that the charge-impurity effect is significantly different from that of the neutral non-magnetic in-plane impurity,
namely,
 for a charge-impurity, MFs indeed exist in the topological superconducting system. It is revealed that a pair of  MFs are bounded by the charge-impurity. As a result, one may manipulate the MFs through controlling the charged impurities, such that they may
 have potential applications in topological quantum computation.

The article is organized as follows. In Sec. II, we introduce
the model and work out the formalism. In Sec. III, we
perform numerical calculations and discuss the obtained
results. Finally, we give a brief summary in Sec. IV.

\section{HAMILTONIAN AND FORMALISM}

We start from a model Hamiltonian that includes the spin-orbital coupling, the Zeeman field, the $s$-wave pairing term, and an additional charged impurity term, which is given by
\begin{eqnarray}
H=-&\sum_{\bf i}[\psi^{\dagger}_{\bf i}({t{\sigma_0}-i
\lambda{\sigma_1}})\psi_{{\bf i}+\hat{x}}\nonumber\\&+\psi^{\dagger}_{\bf i}({t{\sigma_0}-i\lambda{\sigma_2}})\psi_{{\bf i}+\hat{y}}+h.c.]\nonumber\\
&+\sum_{\bf i} \psi^{\dagger}_{\bf i}[(U_{\bf i}-\mu)\sigma_0+h\sigma_3]\psi_{\bf i}\nonumber\\
&+\sum_{\bf i}(\Delta_{\bf i} \psi^\mathrm{T}_{\bf i} i\sigma_2 \psi_{\bf i}+h.c.),
\end{eqnarray}
where $\psi_{\bf i}=(\psi_{{\bf i}\uparrow},\psi_{{\bf
i}\downarrow})^\mathrm{T}$, $\sigma_\nu$ are the identity ($\nu=0$)
and Pauli matrix $(\nu=1,2,3)$, respectively. $\lambda$ and $h$ are the
spin-orbital coupling strength and the effective Zeeman
field, respectively. $U_{\bf i}$ represents the charged impurity potential.

\begin{figure}
\centering
  \includegraphics[width=2.7in]{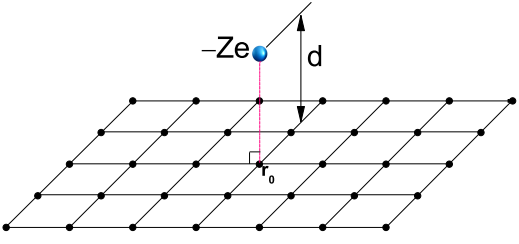}
\caption{(Color online) Schematical illustration of an off-plane charged impurity, with $d$ being the distance between the plane and the impurity. }
\end{figure}

The above Hamiltonian can be diagonalized by solving the Bogoliubov-de
Gennes (BdG) equations,
\begin{equation}
\sum_{\bf j}\left(
\begin{array}{cc}
 H_{{\bf ij}} & \Delta_{{\bf j}}\sigma_3  \\
 \Delta^{*}_{{\bf j}}\sigma_3 & -\sigma_2 H^{*}_{{\bf ij}} {\sigma_2}
\end{array}
\right)  \begin{array}{c} \Psi^{\eta}_{\bf j}
\end{array}
 =E_\eta \begin{array}{c}\Psi^{\eta}_{\bf j}
\end{array},
\end{equation}
with $\Psi^{\eta}_{\bf j}= (u^{\eta}_{{\bf j}\uparrow},u^{\eta}_{{\bf j}\downarrow},v^{\eta}_{{\bf
j}\downarrow},v^{\eta}_{{\bf j}\uparrow})^{\mathrm{T}}$.
The site-dependent order parameter $\Delta_{\bf j}$ is calculated
self-consistently,
\begin{eqnarray}
\Delta_{\bf j}=\frac{V}{2}\sum_\eta u^{\eta}_{{\bf
j}\uparrow}v^{\eta*}_{{\bf j}\downarrow}\tanh (\frac{E_\eta}{2K_B T})
\end{eqnarray}
with $V$ being the pairing strength.

The on-site particle number $n_{\bf i}$ is expressed as,
\begin{equation}
n_{\bf i}=\sum_{\eta\sigma}\mid u^{\eta}_{{\bf i}\sigma} \mid^2 f(E_\eta),
\end{equation}
with $f(E_\eta)$ is the Fermi distribution function.

The local density of states (LDOS) can be calculated as
\begin{equation}
\rho_{\bf i}(\omega)=\sum_\eta [\mid u^{\eta}_{{\bf i}\uparrow} \mid^2
\delta(E_\eta-\omega)+\mid v^{\eta}_{{\bf i}\downarrow} \mid^2
\delta(E_\eta+\omega)],
\end{equation}
where the delta function $\delta(E)$ is taken as
$\delta=\Gamma/[\pi(E^2+\Gamma^2)$], with the quasiparticle damping
$\Gamma=0.01$.

\begin{figure}
\centering
  \includegraphics[width=2.7in]{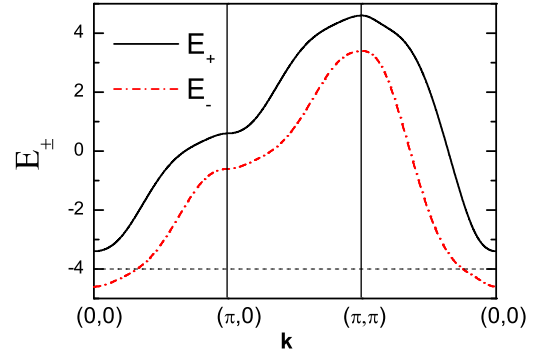}
\caption{(Color online) The renormalized normal state band dispersion in the momentum space [Eq.(6)].}
\end{figure}

In the present work, we consider an off-plane charged impurity carrying the negative electric charge $-Ze$, as sketched in Fig.~1. A repulsive potential  $U_{\bf i}=U_0/\sqrt{({\bf R_i}-{\bf r_0})^2+(d/a)^2}$ is induced by the impurity, with $U_0=Ze^2/(4\pi\varepsilon_0 a)$ ($a$ is the lattice constant). In the following calculations, we set $Z=1$ and $a=4$ {\AA}.  $U_0$ is estimated to be about 3.6 eV.
We use the hopping constant $t$ to be the energy unit and set $U_0=6$. The distance $d$ plays a key role to control the effective scope of the Coulomb potential. As $d$ tends to zero, the potential at ${\bf r_0}$ approaches to the infinite. In this case, the charge-impurity is equivalent to a unitary in-plane neutral non-magnetic impurity~\cite{note}. As $d$ increases, the long range Coulomb interactions take effect. In the present work, we set $d=a$ for illustration. Generally, the MF modes are quite robust for larger $U_0$ and $d$.
The numerical calculations are performed on the quasi-one-dimensional system with the lattice size $400\times 3$, and on the two-dimensional system with the lattice size $48\times 48$.
Note that for both systems, we have verified numerically that no zero energy state exists for a single point-type non-magnetic impurity (not presented here).

The other parameters and the corresponding phase diagram in the absence of impurity have already been discussed in some details~\cite{tzhou}. Here they
are set as $\mu=-4$, $\lambda=0.5$, $h=0.6$, and $V=5$, such that the system is in the topological superconducting phase. We have also checked numerically our main results presented below are not sensitive to the parameters in the topological phase region.
To see the origin of the topological feature more clearly, we transform the bare Hamiltonian into the momentum space, with the two renormalized normal state energy bands $E_{\pm}$, expressed as,
\begin{equation}
E_{\pm}=\varepsilon_{\bf k}\pm\sqrt{h^2+4\lambda^2(\sin^2 k_x+\sin^2 k_y)}.
\end{equation}
Here $\varepsilon_{\bf k}=-2t(\cos k_x+\cos k_y)-\mu$. We plot the above band structure with $\mu=0$ in Fig.~2. Since these two bands are both the superposition of the spin up and spin down electrons. As a result,
with an additional $s$-wave pairing, the Hamiltonian is equivalent to a two-band $p\pm ip$-pairing superconducting system with the opposite chirality, which is usually a topologically trivial superconductor. On the other hand, as is seen in Fig.~2, the two energy bands are separated by the Zeeman field with an energy gap 2$h$ at the $(0,0)$ point. If we set the chemical potential $\mu$ to be inside the gap (e.g., for the case of $\mu=-4$), then the upper band is unoccupied and only the lower band takes effect.
In this case, the system is topologically nontrivial and equivalent to a one-band $p+ip$ superconducting system.

\begin{figure}
\centering
  \includegraphics[width=3in]{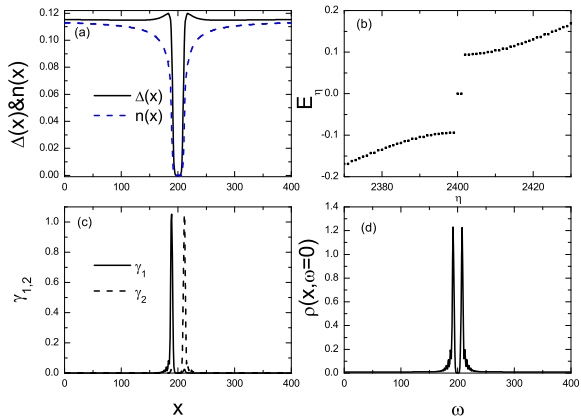}
\caption{(Color online) The numerical results for a $400\times3$ lattice. (a) The self-consistent results of the order parameter $\Delta$ and the site-dependent particle number $n$. (b) The eigenvalues of the Hamiltonian. (c) The spatial distributions of the two MFs. (d) The intensity plot of the LDOS at the zero energy. }
\end{figure}

\section{RESULTS AND DISCUSSION}

We now present the numerical results for a quasi-one-dimensional system. The periodic boundary condition is considered for both $x$-direction and $y$-direction.
For the quasi-one-dimensional lattice, the $y$-dependence of the physical quantities is not important. Thus we here just consider the $x$-dependent of the physical quantities, defined as $A(x)=1/N_y \sum_y A(x,y)$.
 Fig.~3(a) displays the self-consistent results of the superconducting gap and the
site-dependent particle number. As is seen, both the superconducting gap and the on-site particle number are suppressed to zero near the impurity site.
The energy gap increases and tends to be uniform away from the impurity.
As a result, two gap edges at the sites $x=200\pm10$ are induced by the charge-impurity.
This result is consistent with the energy band dispersion shown in Fig.~2. As is seen, the minimum normal state energy is $-4.6$.
Thus the particle number and the pairing gap will be suppressed to zero as $-\mu+U_{\bf i}>4.6$, which yields the gap edges about 10 lattice site away from the impurity.

The existence of the gap edge is important for the MFs.
We now demonstrate numerically the existence of the MF
states.
The information of the MFs can be obtained by diagonizing the BdG equations [Eq.(2)]. The eigenvalues of the BdG-Hamiltonian are plotted in Fig.~3(b). As is seen, two zero-energy eigenvalues are revealed by the numerical results. They are protected by an energy gap about $0.1$. As is known, the two eigenvalues $\pm E$ come from one physical particle, expressed as $C$ and $C^{\dagger}$, which are eigenvectors of the BdG Hamiltonian. For the case of $E=0$, one physical particle can be separated as two MFs. The particle operators of the two MFs can be obtained by $\gamma_1=(C+C^{\dagger})/\sqrt{2}$ and $\gamma_2=i(C^\dagger-C)/\sqrt{2}$. Then the spatial distribution of the MFs $\gamma_{1,2}$ can be studied numerically.
 As is presented in Fig.~3 (a) and (c), a pair of MFs are separated by the impurity and locate near the two gap-edges. This feature of well separated and localized MFs is rather important for manipulation of them, which can be controlled by varying the position and scattering potential of the impurity, having potential applications in topological quantum computing~\cite{naya}.

\begin{figure}
\centering
  \includegraphics[width=3in]{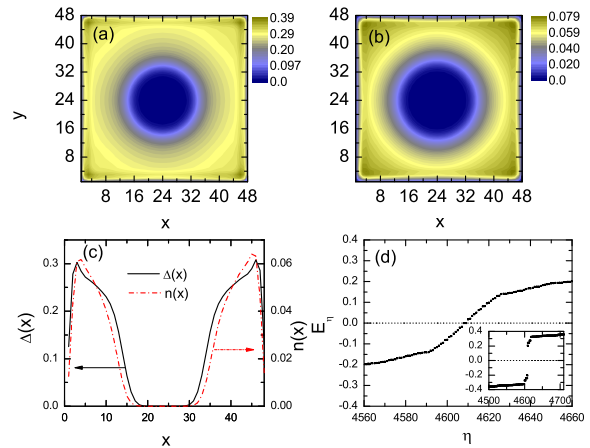}
\caption{(Color online) The numerical results for a $48\times48$ lattice. (a) The intensity plot of the order parameter. (b) The intensity plot of the site-dependent particle number. (c) The two dimensional cuts of the order parameter and the particle number along $x=24$. (d) The eigenvalues of the Hamiltonian the presence of a charge-impurity. Inset of (d): The eigenvalues of the Hamiltonian with the periodic boundary and without the impurity.  }
\end{figure}

The existence of the zero mode can be examined through the LDOS spectra. The intensity plot of the zero energy LDOS at different positions are plotted in Fig.~3(d). As is seen,
there exists a sharp peak at the sites $x=200\pm 10$. It is clear that the zero-energy LDOS is qualitatively the same with the spatial distribution of the two MFs.
 The zero bias peak of the LDOS can be tested by scanning tunnelling microscope experiments. Thus the indication of MFs may be tested experimentally.

Let us turn to present the numerical results of a two-dimensional lattice with a charge-impurity located at ${\bf r_0}=(24,24)$. The intensity plot of the order parameter magnitude and the on-site particle number with the open boundary condition are displayed in Figs.~4(a) and 4(b), respectively. Their two-dimensional cuts (along $y=24$) are plotted in Fig.~4(c). As is seen, both the order parameter magnitude and the particle number are near zero as $\mid {\bf R_i}-{\bf r_0}\mid <10$. This result is similar to the case of quasi-one-dimensional system and can be understood through the band structure shown in Fig.~2.
As $\mid {\bf R_i}-{\bf r_0}\mid >10$, the effective on-site potentials cross the lower energy band, and thus the on-site particle number and the gap magnitude increase. As a result, the gap-edges at the sites $\mid {\bf R_i}-{\bf r_0}\mid =10$ form. An edge state should exist and the MFs may exist near the gap-edges.
The eigenvalues of the two-dimensional Hamiltonian is plotted in Fig.~4(d). We also plot the eigenvalues for the case of the uniform pairing states with periodic boundary in the inset of Fig.~4(d).
For the two-dimensional lattice in the presence of the boundaries, it is expected that an edge state crossing the zero energy should exist.
 In the uniform superconducting state, an energy gap about 0.2 is seen clearly. In the presence of a charge-impurity, as seen in Fig.~4(d), the edge state crossing the zero energy exists, protected by a bulk energy gap about 0.2. Two zero-energy eigenvalues can be seen clearly, related to the MF modes near the gap edges.

Similar to the case of the quasi-one-dimensional lattice, the two MFs can be studied numerically from the eigenvectors of the zero eigenvalues. The numerical results of the spatial distribution of the two MF states are presented in Figs.~5(a) and 5(b). We also plotted the zero energy LDOS in Fig.~5(c) to disclose possible experimental observation for
the existence and distribution of the zero
mode. Their two dimensional cuts along the line $y=24$ are plotted in Fig.~5(d).

\begin{figure}
\centering
  \includegraphics[width=3in]{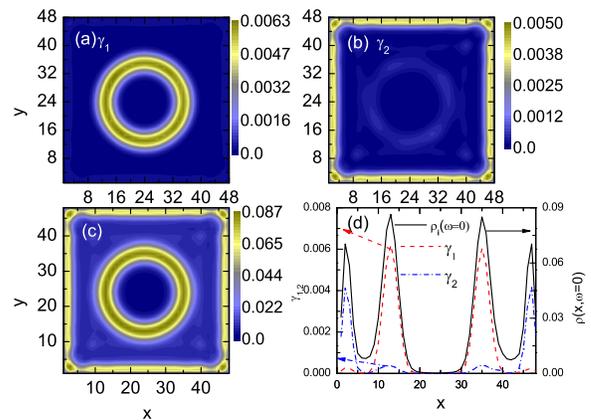}
\caption{(Color online) The numerical results for the $48\times48$ lattice. (a) The spatial distribution of $\gamma_1$. (b) The spatial distribution of $\gamma_2$. (c) The intensity plot of the zero energy LDOS. (d) The two dimensional cults of panels (a-c) along $y=24$. }
\end{figure}

As is shown in Figs.~5(a) and 5(b), the two well separated MFs are identified numerically. The spatial distribution of $\gamma_1$ forms a circle with the radius about 10 lattice sizes. Thus it appears near the superconductor-insulator boundary. Another MF $\gamma_2$ appears at the system boundary. These results are qualitatively the same as those in the case of the vortex-induced MFs in topological superconductors~\cite{read}. In this sense, here the charge-impurity may be viewed as an artificially created vortex. One may manipulate the MFs through controlling the charge-impurity.
Similar to the case of the quasi-one-dimensional lattice, the zero-mode can be detected experimentally through the zero-energy LDOS. The intensity plot of the zero-energy LDOS is displayed in Fig.~5(d). Here the spectrum is qualitatively consistent with the
distributions of the two MF states. As a result, a signature of MFs
may also be obtained through the LDOS spectra for the two-dimensional lattice system.

Finally, we would like to remark the significances of the present work. First, we here propose an effective tool to realize the MFs subject to an off-plane charge-impurity. The MFs are bounded by the impurity. Technically they may be well controlled through operating the impurities. Secondly, the two MFs are well separated in space and there is nearly no overlap. This is different from the case of the MF states induced by a harmonic potential~\cite{zhou}.
Actually, the harmonic potential increases and tends to be infinite far away from the trapping center. Thus the superconducting region will only exist in a small region. Thus the two MFs subject to a harmonic potential are close in space. As a result, the two MFs should overlap in space. On the other hand, for the MFs induced by the Coulomb interaction, the potential tends to zero far away from the impurity center. For the large system size, the two MFs are sufficiently separated. This feature is of importance and may merit further application in topological quantum computation. Thirdly,
for the two-dimensional lattice, the property of the MFs is qualitatively similar to that of the vortex-bounded MFs. Therefore, the non-Abelian
statistics feature may be seen more easily in experiments as the charge-impurities can be well-controlled,  which is of fundamental interest. At last, it is worthwhile pointing out that there exists another significant difference between the present work and the previous one on the MFs induced by a harmonic potential~\cite{zhou}. In Ref.~\cite{zhou}, the numerical calculation is based on a typical model considering a spin-dependent hopping term, which was proposed to be realized in cold atom systems.
While here our numerical calculations and main results are based on a standard model of topological superconductors.
They
may be generalized to other topological superconducting systems, e.g., the $p+ip$ superconducting systems or the semiconductor nonowire/s-wave superconductor
heterostructure system. 

\section{summary}
In summary, we have studied numerically the effect of an off-plane charged impurity in topological superconductors. We have revealed that the MFs can be induced by the impurity. For a quasi-one dimensional system, a pair of MFs locate at the two sides of the impurity, while for a two dimensional system, one MF is bounded by the impurity and the other appears at the boundary of the system. The LDOS spectra have also been calculated, based on which a clear indication of the MFs may be observed experimentally.

This work was supported
by the NSFC (Grant No. 11374005 and No. 11204138), the NCET (Grant No. NCET-12-0626), Jiangsu Qingnan engineering project, the GRF (Grant Nos. HKU7045/13P and and HKU173051/14P) and the CRF (Grant No. HKU8/11G) of Hong Kong.


\begin{thebibliography}{99}
\bibitem{note} For the in-plane charged impurity, the long range interactions are perfectly screened. An additional screened potential is required to simulate the real system.
\end{thebibliography}
\end{document}